\documentclass[a4paper,11pt]{article}

\usepackage{amsmath}
\usepackage{amssymb}
\usepackage{cite}
\usepackage{epsfig}
\usepackage{cancel}
\usepackage{feynmp}
\usepackage{xspace}
\usepackage{flafter}

\def\spa#1.#2{\left\langle#1\,#2\right\rangle}
\def\spb#1.#2{\left[#1\,#2\right]}
\def\spaa#1.#2.#3{\langle\mskip-1mu{#1}
                  | #2 | {#3}\mskip-1mu\rangle}
\def\spbb#1.#2.#3{[\mskip-1mu{#1}
                  | #2 | {#3}\mskip-1mu]}
\def\spab#1.#2.#3{\langle\mskip-1mu{#1}
                  | #2 | {#3}\mskip-1mu\rangle}
\def\spba#1.#2.#3{\langle\mskip-1mu{#1}^+
                  | #2 | {#3}^+\mskip-1mu\rangle}
\def\spav#1.#2.#3{\|\mskip-1mu{#1}
                  | #2 | {#3}\mskip-1mu\|^2}
\def\jc#1.#2.#3{j^{#1}_{#2#3}}
\newcommand{\Ca}{\ensuremath{C_{\!A}}\xspace}
\newcommand{\Cf}{\ensuremath{C_{\!F}}\xspace}
\newcommand{\Nc}{\ensuremath{N_{\!C}}\xspace}

\newcommand{\as}{\ensuremath{\alpha_s}\xspace}

\title{
\begin{normalsize}
\begin{flushright}
CERN-PH-TH/2009-202\\
\end{flushright}
\end{normalsize}
\vspace*{1cm}
The Factorisation of the $t$-channel Pole\\in Quark-Gluon Scattering}
\author{Jeppe~R.~Andersen$^{a}$,
  Jennifer~M.~Smillie$^{b}$\\\mbox{}\\$^a$Theory Division, Physics
  Department, CERN, CH-1211 Geneva 23, Switzerland\\$^b$ Department of Physics and
  Astronomy, UCL, Gower Street, London, WC1E 6BT.}

\begin{document}
\maketitle
\begin{abstract}
  By exploring the scattering of specific helicity states in quark-gluon
  scattering at tree level we show explicitly that the $t$-channel pole can
  be described exactly as a contraction of two local currents. Furthermore,
  we demonstrate that out of eight non-zero helicity possibilities, only two
  suppressed channels have contributions that are not pure, factorised
  $t$-channel poles. We thereby extract a gauge-invariant definition for the
  $t$-channel current generated by the scattering of a gluon. This offers a
  slight improvement in the description of gluon scattering in the framework
  of Ref.\cite{Andersen:2009nu} for the prediction of $n$-jet rates at hadron
  colliders.
\end{abstract}

\section{Introduction}
\label{sec:introduction}

Most studies at the LHC of physics from both within and possibly beyond the
Standard Model will require a detailed understanding of not just the rate but
also the topology of hard multi-jet events. The vast phase space opened by
the centre-of-mass energy of the accelerator can counter-act the \as-suppression
of further radiation in the hard scattering matrix element\footnote{This is
  true in particular for processes where the partonic cross section is not
  suppressed with increasing partonic centre-of-mass energy $\hat s$, such as
  e.g.~$2\!\to\!2$-processes which can proceed through a $t$-channel exchange
  of a gluon. For large $\hat s$ the partonic cross section for such
  processes limits to a constant depending on the transverse momentum
  only. All other $2\!\to\!2$ processes are suppressed by powers of $\hat
  s$.}.


This means that it is not only relevant to calculate processes of
ever higher multiplicity at the lowest order in perturbation theory, but the
description of even hard radiative corrections to these tree-level
configurations become increasingly relevant. To date, the radiative corrections
for LHC processes with two or more QCD charged particles in the final state
is known in full fixed order perturbation theory only to the first order
(i.e.~the process is known to next-to-leading order).  Radiative corrections
beyond the first order have traditionally been approximated within a parton
shower-approach\cite{Bahr:2008pv,Sjostrand:2007gs,Gleisberg:2008ta}. The
approximations applied to the real emissions become exact in the soft and
collinear limits, and result in a sufficiently simple formalism that
all-order results can be obtained. Virtual corrections are defined by keeping
the shower evolution unitary (i.e.~the probability for emitting or not
emitting equal to one; in the language of fixed order calculations, the
$K$-factor induced by the parton
shower for the inclusive cross section is one).

The perturbative corrections simplify not just in the soft and collinear
limit, but also in the limit of large invariant mass between all produced
particles, the limit of so-called \emph{Multi-Regge-Kinematics} (MRK), where
the all-order perturbative results for $2\to n$-scattering are known not just
for the real but also virtual
corrections\cite{Fadin:1975cb,Kuraev:1976ge,Kuraev:1977fs}. This limit is of
interest when the focus is on describing correctly the number and topology of
jets, rather than the radiation within each jet, since any jet-definition
introduces a requirement of a non-negligible invariant mass between the
constituents of separate jets. 

\begin{figure}[!hbtp]
  \centering
  
  \vspace{0.5cm}
  \input{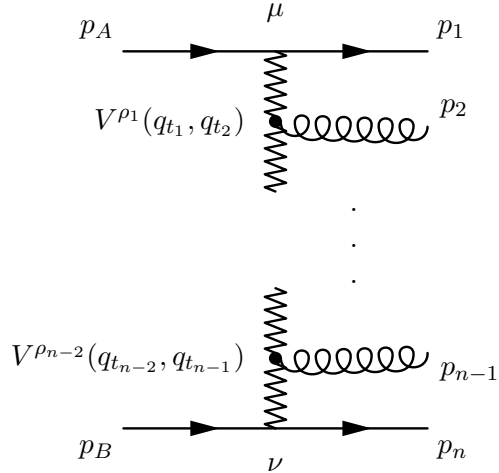}
  
  \vspace{0.5cm}
  \caption{This figure illustrates the analytic structure of the
    approximating amplitudes. The $2\!\to\!n$ scattering amplitude is
    described by a basic $2\!\to\!2$ process of current contractions under a
    $t$-channel exchange, with effective vertices describing the effect of
    additional gluon radiation. This ensures the correct MRK limit.}
  \label{fig:structure}
\end{figure}

We have recently presented a framework\cite{Andersen:2009nu}, which not only
reproduces the exact $2\to n$ results in the MRK limit, but also reproduces
to a good degree the results obtained using full perturbative QCD
order-by-order (for the low orders where such results can be obtained) for
completely inclusive calculations, i.e.~without special cuts in phase
space. The amplitude for the scattering is described by a basic $2\to
2$-scattering under the exchange of the current generated by the deflection
of each particle, supplemented by effective vertices for the extra gluon
emission. These effective vertices take into account the emission off each of
the four legs of the basic (or backbone) $2\to2$ process, and emission off
the exchanged current. The formulation of Ref.\cite{Andersen:2009nu} in terms
of current scattering of specific helicity states provides the crucial
improvement over initial efforts in
Ref.\cite{Andersen:2008ue,Andersen:2008gc}, and extends the phase space
region of applicability even further. Fig.~\ref{fig:structure} illustrates
the resulting structure for the approximation for the scattering amplitude in
terms of the contraction of $t$-channel currents with the two incoming
particles, and the generation of additional emission by effective
vertices. Rapidity ordering is implied as $y_1>y_i>y_n$. The diagram on the
figure is \emph{not} a Feynman diagram, but is a one-to-one representation of
the formula for the approximation of the matrix elements. The point of
constructing the approximations to the perturbative series is to obtain a
formalism which is sufficiently simple to allow for the all-order sum to be
constructed directly, while being sufficiently accurate when compared order by order
to the full fixed order results where these can be obtained. See
Ref.\cite{Andersen:2009nu} and Section~\ref{sec:results} for further details.

In the MRK limit, the kinematical dependence of the amplitude for
quark-quark, quark-gluon and gluon-gluon scattering is identical, and the
scattering amplitude differs only by colour factors. In this limit, the
scattering amplitude is dominated by the behaviour dictated by the poles in
the $t$-channel momenta. Therefore, the picture advocated in
Ref.\cite{Andersen:2009nu} is built on the basic structure of the scattering
of two different quark flavours $qQ\to qQ$, which at lowest order proceeds
through the exchange of a single gluon in the $t$-channel, with the gluon
channels adjusted just by colour factors. 

The description of the basic $2\!\to\!2$-scattering is therefore one of the contraction of
two generated currents, each of the form $A^\mu=\overline\psi\gamma^\mu\psi$. We will call
this form of the matrix element ``factorised'', since each current obviously depends on
the momenta of the local scattering spinors only. As already mentioned, the factorised
form arising for the scattering of quarks was used also for gluon scattering in
Ref.\cite{Andersen:2009nu}, changing only the effective colour factor. This results in the
right MRK limit also for processes with gluon scattering.

A priori, one might worry about extending the simple description in
quark-quark scattering to processes involving gluons, since e.g.~there are
three Feynman diagrams contributing to $qg\to qg$ instead of the one in
$qQ\to qQ$, with apparent singularities in the $s$ and $u$-channels (see
Fig.~\ref{fig:standu}). In Sec.~\ref{sec:gluon-amplitudes} we will show
explicitly how the full tree-level scattering for $qg\to qg$ obeys complete
factorisation according to the above definition for all except two (suppressed)
out of eight helicity configurations. Even in these two suppressed channels,
the $t$-channel singularity is completely factorised. We thereby obtain a
gauge-invariant definition of the off-shell current generated by the
scattering of a gluon, by using the natural definition of the current as
the full coefficient of the $t$-channel pole. We can thereby define an
improved \emph{impact factor} for the gluon, which ensures that the
description of $qg\to qg$ is exact (for 6 out of 8 helicity configurations,
and for all 4 dominant ones). In Sec.~\ref{sec:results} we will investigate
the slight improvement of the results in Ref.\cite{Andersen:2009nu} on the
description of jet production at the LHC offered by the inclusion of these
sub-leading corrections.

\section{Quark-Gluon Amplitudes}
\label{sec:gluon-amplitudes}

\begin{figure}[!btp]
  \centering
  \input{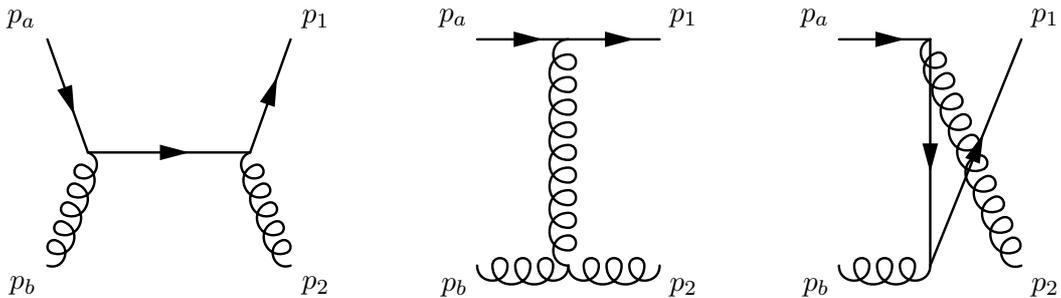}
  \caption{The $s$-, $t$- and $u$-channel processes which contribute to $q^-(p_a) +
    g^+(p_b) \to q^-(p_1) + g^+(p_2)$.}
  \label{fig:standu}
\end{figure}

We start with the $2\to 2$ process $qg \to qg$ (Fig.~\ref{fig:standu}) and consider the
different helicity contributions, beginning with $q^-(p_a) + g^+(p_b) \to q^-(p_1) +
g^+(p_2)$.  We make the following gauge choice for the polarisation vectors:
\begin{align}
  \label{eq:gauge}
  \begin{split}
     &\varepsilon_{2\sigma}^{+*}=\frac{\bar u_b^- \gamma_\sigma u_2^-}{\sqrt{2}\ \bar u_b^-
       u_2^+} \qquad \varepsilon_{2\sigma}^{-*}=\frac{-\bar u_2^- \gamma_\sigma
       u_b^-}{\sqrt{2} \bar u_b^+ u_2^-} \\
     \varepsilon_{b\rho}^+ = &\frac{-\bar u_2^+ \gamma_\rho u_b^+}{\sqrt{2}\ \bar
       u_2^+ u_b^-}=\frac{-\bar u_b^- \gamma_\rho u_2^-}{\sqrt{2}\ \bar u_2^+ u_b^-}
     \qquad \varepsilon_{b\rho}^- = \frac{\bar u_2^- \gamma_\rho u_b^-}{\sqrt{2} \bar
       u_2^- u_b^+}. 
  \end{split}
\end{align}
This particular choice gives a symmetric form, and keeps the
factorisation explicit between forward moving particles ($p_a,p_1$) and backward moving
particles ($p_b,p_2$).  Using the conventions outlined in Appendix \ref{sec:spin-repr} and
the following shorthands:
\begin{align}
  \label{eq:shorthands}
  \spab{i}.\mu.{j}=\bar u_i^- \gamma^\mu u_j^-, \quad \spa{i}.{j}=\bar u_i^- u_j^+, \quad
  \mathrm{and} \quad \spb{i}.{j}=\bar u_i^+ u_j^-,
\end{align}
we get (where $A_x$ is the amplitude for the $x-$channel diagram):
\begin{eqnarray}
  \label{eq:amps}
  A_s&=& \frac{g^2 t^b_{ae} t^2_{e1} }{\hat t}\ \times\
  -\sqrt{\frac{p_2^-}{p_b^-}} \frac{p_{2\perp}^*}{|p_{2\perp}|}
  \spab{b}.\sigma.2 \ \times\ \spab1.\sigma.a. \nonumber \\
  A_t&=& 0 .\\  \nonumber 
  A_u&=& \frac{g^2 t^2_{ae}t^b_{e1}}{\hat t}\ \times \ \sqrt{\frac{p_b^-}{p_2^-}}
  \frac{p_{2\perp}^*}{|p_{2\perp}|} \spab{b}.\rho.2\ \times \ \spab1.\rho.a.
\end{eqnarray}
This gives a sum of
\begin{equation}
  \label{eq:sum}
  \frac{g^2}{\hat t}\times \frac{p_{2\perp}^*}{|p_{2\perp}|} \left( t^2_{ae}t^b_{e1}
    \sqrt{\frac{p_b^-}{p_2^-}} - t^b_{ae} t^2_{e1} \sqrt{\frac{p_2^-}{p_b^-}} \right)
  \spab{b}.\sigma.2\ \times \spab1.\sigma.a.
\end{equation}
In the HE limit, $p_b^- \sim p_2^-$ so the sum is
\begin{equation}
  \label{eq:FINAL}
  \frac{g^2}{\hat t}\ i f^{2bm}t^m_{a1}\ \times \
  \frac{p_{2\perp}^*}{|p_{2\perp}|}\spab{b}.\sigma.2\ \times \spab1.\sigma.a
\end{equation}
which agrees (up to an irrelevant phase) with the structure used in
\cite{Andersen:2009nu}.  However, the crucial result in Eq.~\eqref{eq:sum} is
that this helicity amplitude for quark-gluon scattering is still expressible
exactly as a scattering under exchange of a $t$-channel gluon
current. However, the current generated by the scattering of a gluon is
slightly more complicated (by the terms in the brackets of
Eq.~\eqref{eq:sum}) than that generated by a quark.  The colour summed and
averaged scattering matrix element is
\begin{align}
  \label{eq:sqcol}
  |M^t_{q^-g^+\to q^- g^+}|^2\ =  \frac{g^4}{\hat t_{a1} \hat t_{b2}}\frac{C_F} {N_c^2-1}\
  \left( \frac 1 2 \frac{p_b^{-2}+p_2^{-2}}{p_b^- p_2^-}\ \left(C_A -\frac 1
      {C_A}\right)+ \frac{1}{C_A}\right)\ | \spab{b}.\rho.2 \spab1.\rho.a |^2.
\end{align}
In this case, $\hat t_{a1}=\hat t_{b2}$, but we write it this way in anticipation of the
multijet case.  Cast in this form, we see directly that this helicity scattering of quarks
and gluons is identical to that of the scattering of two different quark flavours with a
replacement of \Cf by the colour factor
\begin{align}
  \label{eq:finalcol}
  \frac 1 2 \left(C_A-\frac 1
    {C_A}\right)\left(\frac{p_b^-}{p_2^-}+\frac{p_2^-}{p_b^-}\right) + \frac
  1 {C_A}.
\end{align}
We note that in the MRK limit ($p_b^-\to p_2^-$), this tends to \Ca, as used
in Ref.\cite{Andersen:2009nu}. Eq.~\eqref{eq:finalcol} expresses how the
strength of the current increases with increasing acceleration of the
scattering gluon (as $\left(\frac{p_b^-}{p_2^-}+\frac{p_2^-}{p_b^-}\right)$
increases). We will therefore call the result of Eq.~\eqref{eq:finalcol} 
the \emph{Colour Acceleration Multiplier} (CAM).

For the same process with positive helicity quarks ($q^+(p_a) + g^+(p_b) \to
q^+(p_1) + g^+(p_2)$), the only difference is that $\spab1.\rho.a$ becomes
$\spab{a}.\rho.1$ which leads to the same gluon impact factor.  The
processes with negative helicity gluons can be found by taking the complex
conjugate of these results, and because the new multiplicative factor is
real, we again find the corresponding quark current multiplied by
Eq.~\eqref{eq:finalcol}.

The amplitudes for the scatterings where the gluon helicity is \emph{not}
flipped all scale as $\hat s/\hat t$ in the MRK limit. The cases where the
gluon flips helicity are more complicated, but are calculated in the same
way. The two distinct cases are $q^- g^- \to q^- g^+$, which gives:
\begin{align}
  \label{eq:allmflip}
  A_s&= \frac{g^2 t^b_{ae} t^2_{e1} }{\hat t }\ \times\
  \left(\sqrt{\frac{p_2^-}{p_b^-}}\frac{p_{2\perp}}{|p_{2\perp}|}
  \spab{b}.\sigma.2 +2 p_b^\sigma \frac{\hat t}{\hat s} \right) \ \times\
\spab1.\sigma.a \nonumber \\  
A_t&= -ig^2t^m_{a1}f^{m2b}\ \times\ \frac{(p_2+p_b)^\mu}{\hat t}\ \times\ \spab1.\mu.a
  \\   \nonumber  
  A_u&=\frac{-g^2 t^2_{ae}t^b_{e1}}{\hat t}\ \times \ \left( \sqrt{\frac{p_b^-}{p_2^-}}
  \frac{p_{2\perp}^*}{|p_{2\perp}|} \spab2.\rho.b +2p_2^\rho \frac{\hat t}{\hat u}
\right)\ \times \ \spab1.\rho.a, 
\end{align}
and $q^- g^+ \to q^- g^-$ which gives:
\begin{align}
  \label{eq:mixflip}
    A_s&= \frac{-g^2 t^b_{ae} t^2_{e1} }{\hat t }\ \times\
  \sqrt{\frac{p_2^-}{p_b^-}} \frac{p_{2\perp}^*}{|p_{2\perp}|}
  \spab2.\sigma.b \ \times\ \spab1.\sigma.a \nonumber \\  
  A_t&= ig^2t^m_{a1}f^{m2b}\ \times\ \frac{(p_2+p_b)^\mu}{\hat t}\ \times\ \spab1.\mu.a
  \\   \nonumber  
  A_u&=\frac{g^2 t^2_{ae}t^b_{e1}}{\hat t}\ \times \  \sqrt{\frac{p_b^-}{p_2^-}}
  \frac{p_{2\perp}}{|p_{2\perp}|} \spab{b}.\rho.2 \ \times \ \spab1.\rho.a. 
\end{align}
We see that only for the helicity configuration where the incoming gluon has the same
helicity as that of the quark \emph{and} the helicity of the gluon is flipped is there a
contribution which is \emph{not} expressible as a current contraction over a $t$-channel
pole. These terms which give rise to the poles in $\hat u$ and $\hat s$ have their origin
in these helicity configurations only. We are interested only in describing the
$t$-channel pole.

The amplitudes with positive helicity quarks can be obtained by complex
conjugation.  We notice that between Eqs.~\eqref{eq:allmflip} and
\eqref{eq:mixflip}, $A_s$ and $A_u$ are swapped (as we would expect).  One
can check explicitly that the amplitudes for the scatterings which flip the
gluon helicity all vanish in the MRK limit.

We now seek to find the equivalent of Eq.~\eqref{eq:sqcol} for the non-helicity
conserving amplitudes.  We use the shorthands
\begin{align}
    \label{eq:helcurrents}
    j^{-,\mu}_{ij}\ =\ \langle j|\mu|i\rangle, \qquad \mathrm{and} \qquad
    j^{+,\mu}_{ij}\ =\ \langle i|\mu|j\rangle.
\end{align}
Then the matrix element squared with summed and averaged colour is
\begin{align}
  \begin{split}
    \label{eq:colsumandaverhelflipamp}
    |M^t_{q^-g^-\to q^- g^+}|^2\ =&\ \frac {g^4}{\hat t_{a1}\ \hat t_{b2}} \frac 1
    {\Ca(\Nc^2-1)}\ \frac{\Cf}2 *\\
    \Bigg(&\left(\Ca^2-1\right) \ \left(\frac{p_2^-}{p_b^-} |\jc+.b.2
      . \jc-.a.1|^2+\frac{p_b^-}{p_2^-}|\jc-.b.2 . \jc-.a.1|^2 )\right)\\
    &\ +2\ \Ca^2\ |(p_2+p_b).\jc-.a.1|^2\\
    &\ +2\Ca^2\ \sqrt{\frac{p_2^-}{p_b^-}}\ \Re\left[\frac{p_{2\perp}^*}{|p_{2\perp}|}\
      \cdot\ (p_2+p_b).\jc-.a.1\ \cdot\ (\jc^+.b.2 . \jc^-.a.1)^*\right]\\
    &\ +2\Ca^2\ \sqrt{\frac{p_b^-}{p_2^-}}\Re\left[\frac{p_{2\perp}}{|p_{2\perp}|}\
      \cdot \
      (p_2+p_b).\jc-.a.1\ \cdot \ (\jc-.b.2 . \jc-.a.1)^*\right]\\
    &\ +2\Re\left[\frac{p_{2\perp}}{p_{2\perp}^*} \jc+.b.2 . \jc-.a.1\ \cdot\ (\jc-.b.2
      . \jc-.a.1)^* \right]\Bigg).
    \end{split}
\end{align}
The result for $|M^t_{q^-g^+\to q^- g^-}|^2$ is very similar, with the $j_{b2}$ currents
reversed and appropriate phases complex conjugated, as can be seen by comparing
Eqs.~\eqref{eq:allmflip} and \eqref{eq:mixflip}.  These matrix elements are more
complicated than Eq.~\eqref{eq:sum}, but are just the sum of terms with a similar form.

In all the helicity configurations discussed so far, the gluon has been taken to be moving
in the backward direction.  We see the same form for all helicity configurations where the
gluon is taken to be moving in the forward direction so do not repeat the results here.
The only difference comes in phases arising from our conventions for the spinors, given in
Appendix~\ref{sec:spin-repr}.

The over-all conclusion of this section is that using scattering of helicity
states, it is possibly to extract a gauge-invariant definition of the
$t$-channel current generated by the deflection of a gluon. Only two out of
eight helicity configurations require an approximation (and these
configurations are suppressed in the relevant limit). The improved
description of the current generated by a gluon can now immediately be
incorporated in the framework of Ref.\cite{Andersen:2009nu} for the
description of also $gg$-scattering, and the leading contribution to
$2\!\to\!n$ scattering processes. 

The case of $2\!\to\!2$ pure helicity-non-flipping gluon scattering is
described as simply the scattering of two quark-generated currents, but with
a colour acceleration multiplier (Eq.~\eqref{eq:finalcol}) for each
current. The possibility of one helicity flipping is then described simply as
Eq.~\eqref{eq:colsumandaverhelflipamp}, but with a CAM for the non-flipping
gluon current. Since the contribution from the single gluon helicity-flipping
amplitudes is small, we refrain from a description of the (double suppressed)
contribution of a flip in the helicity of both scattered gluons.

The square of the $2\!\to\!n$ scattering amplitude is approximated by the sum
over the square of the basic $2\!\to\!2$ current contractions (for each
helicity possibility), multiplied by one (gauge invariant) effective vertex
for each additional gluon emission. See Ref.\cite{Andersen:2009nu} for
further details.

In the next section we will access directly the quality of the approximations
in the description of hard multi-jet production by comparing to the full
tree-level results.


\section{Results}
\label{sec:results}

\begin{figure}[!btp]
  \centering
  \epsfig{width=0.49\textwidth,file=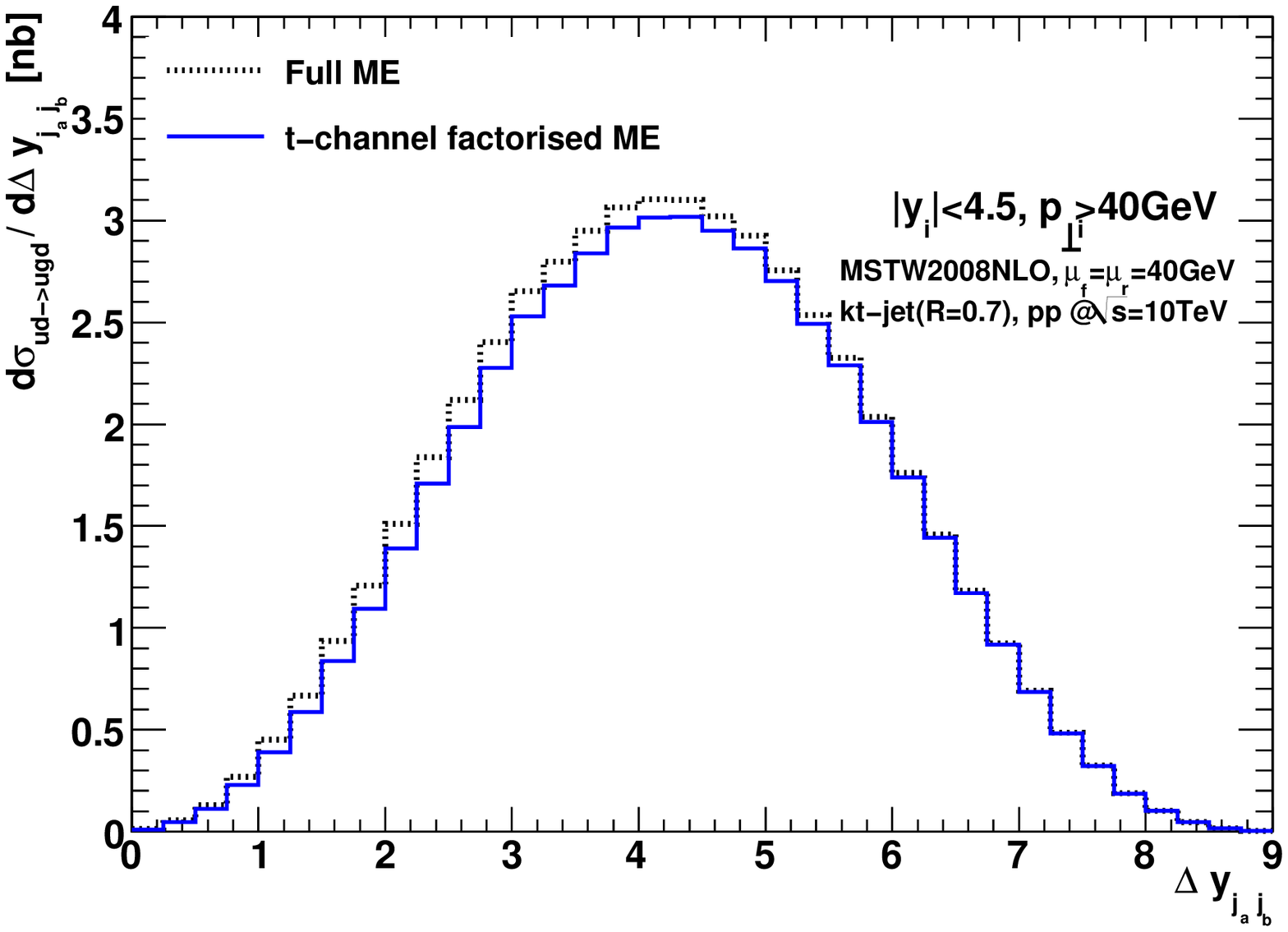}
  \epsfig{width=0.49\textwidth,file=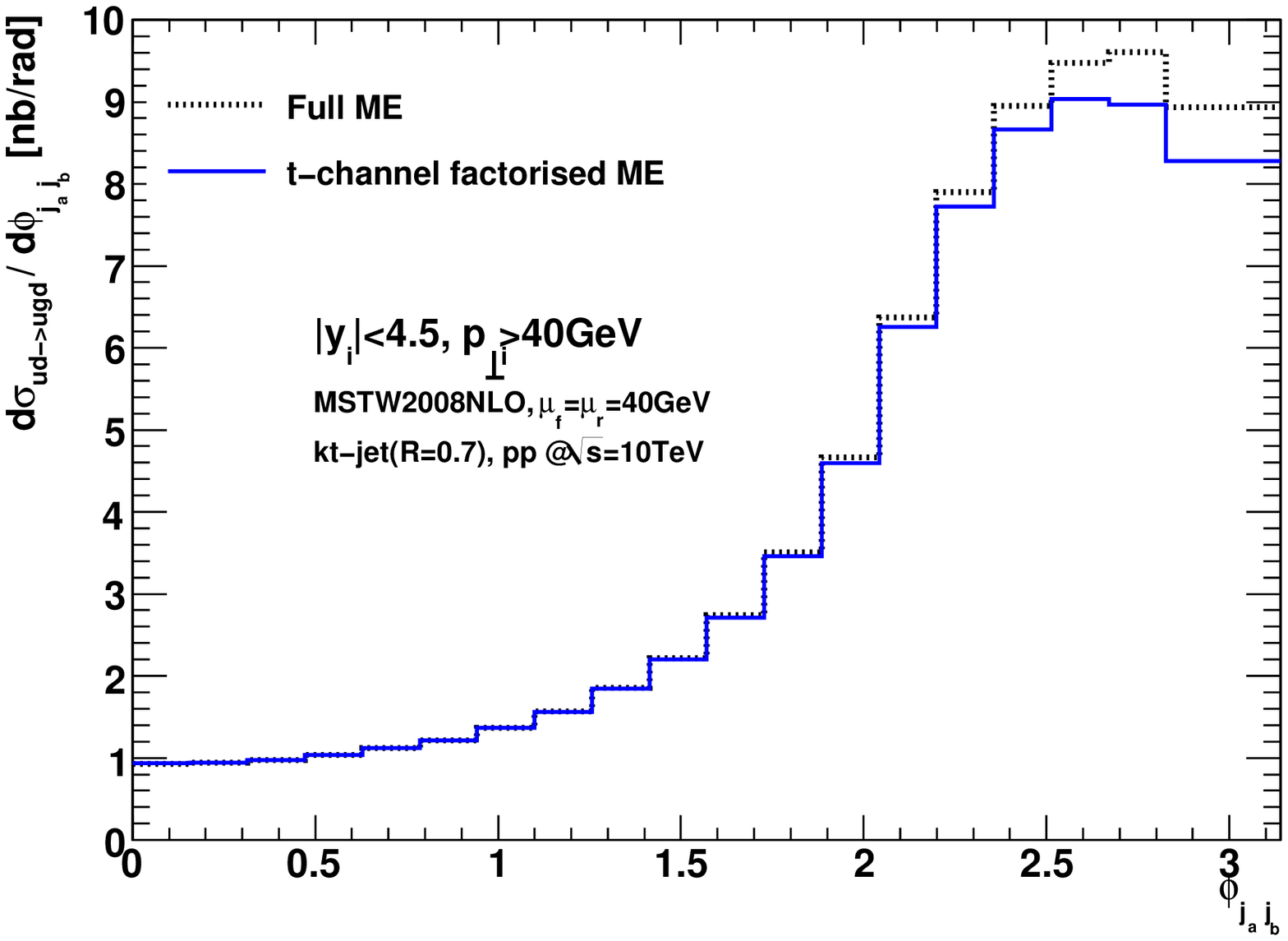} \\
  (a) \hspace{7.2cm}(b)\hspace{0.1cm}\\
  \epsfig{width=0.49\textwidth,file=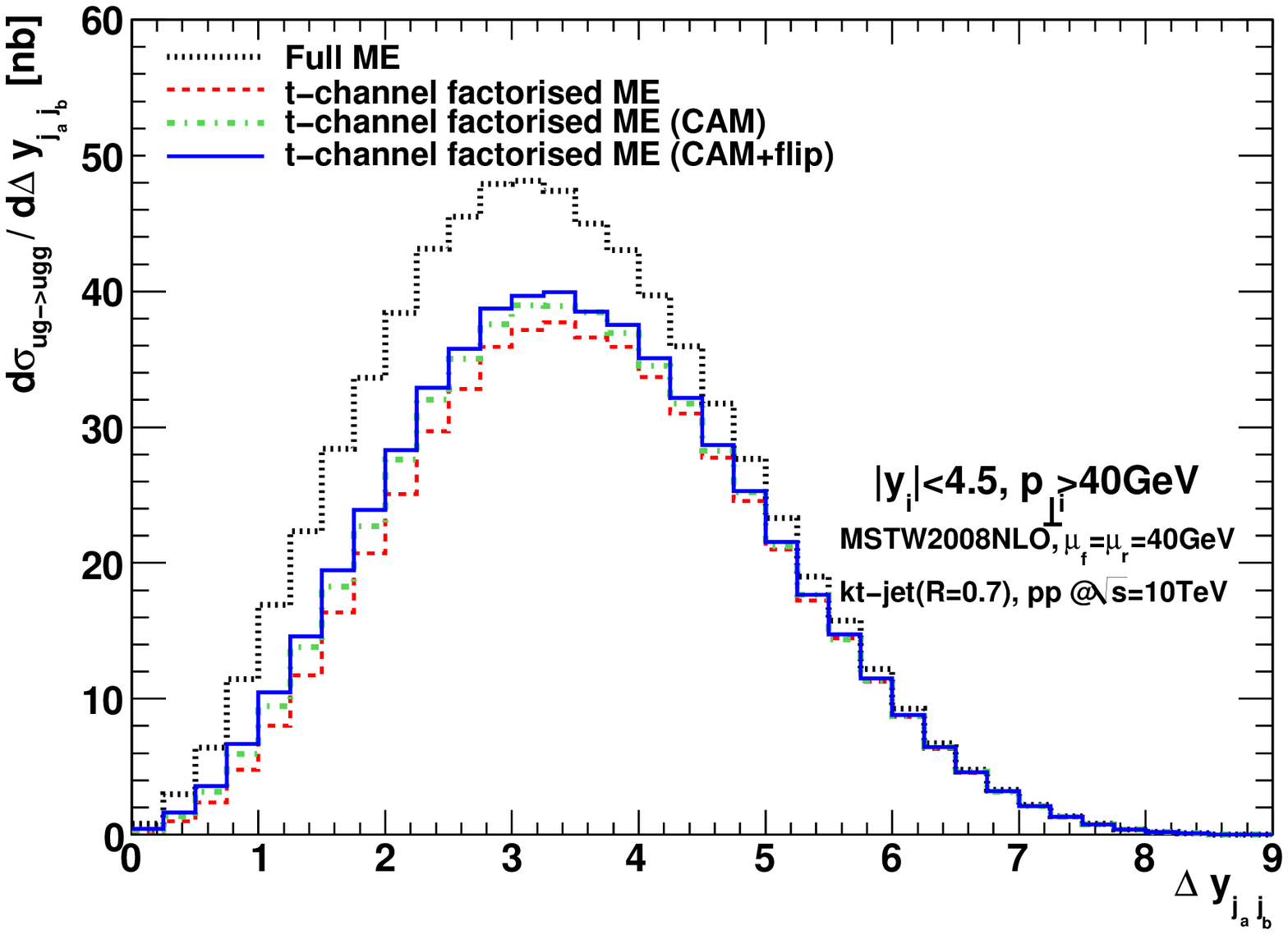}
  \epsfig{width=0.49\textwidth,file=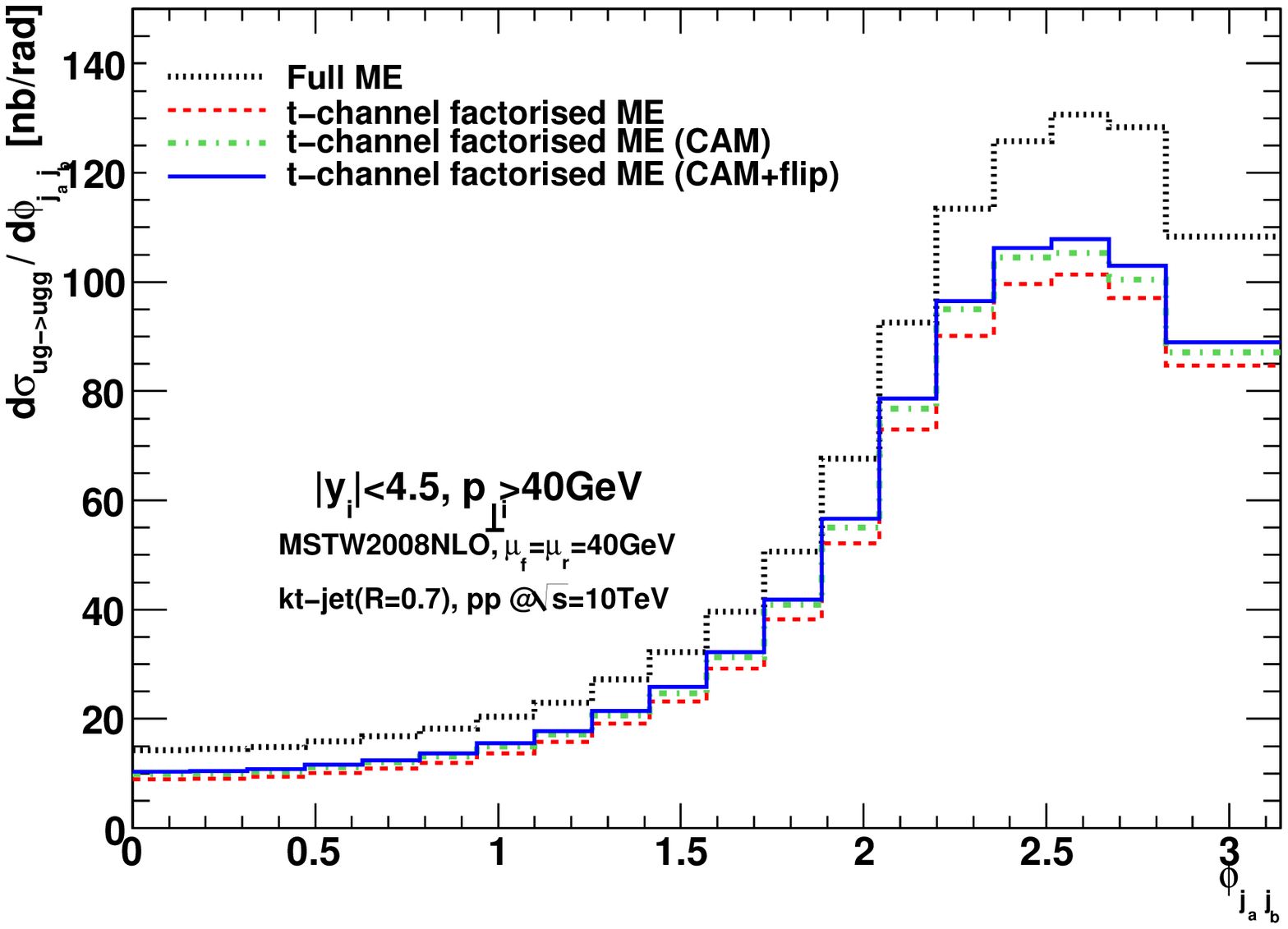} \\
  (c) \hspace{7.2cm}(d)\hspace{0.1cm}\\
  \epsfig{width=0.49\textwidth,file=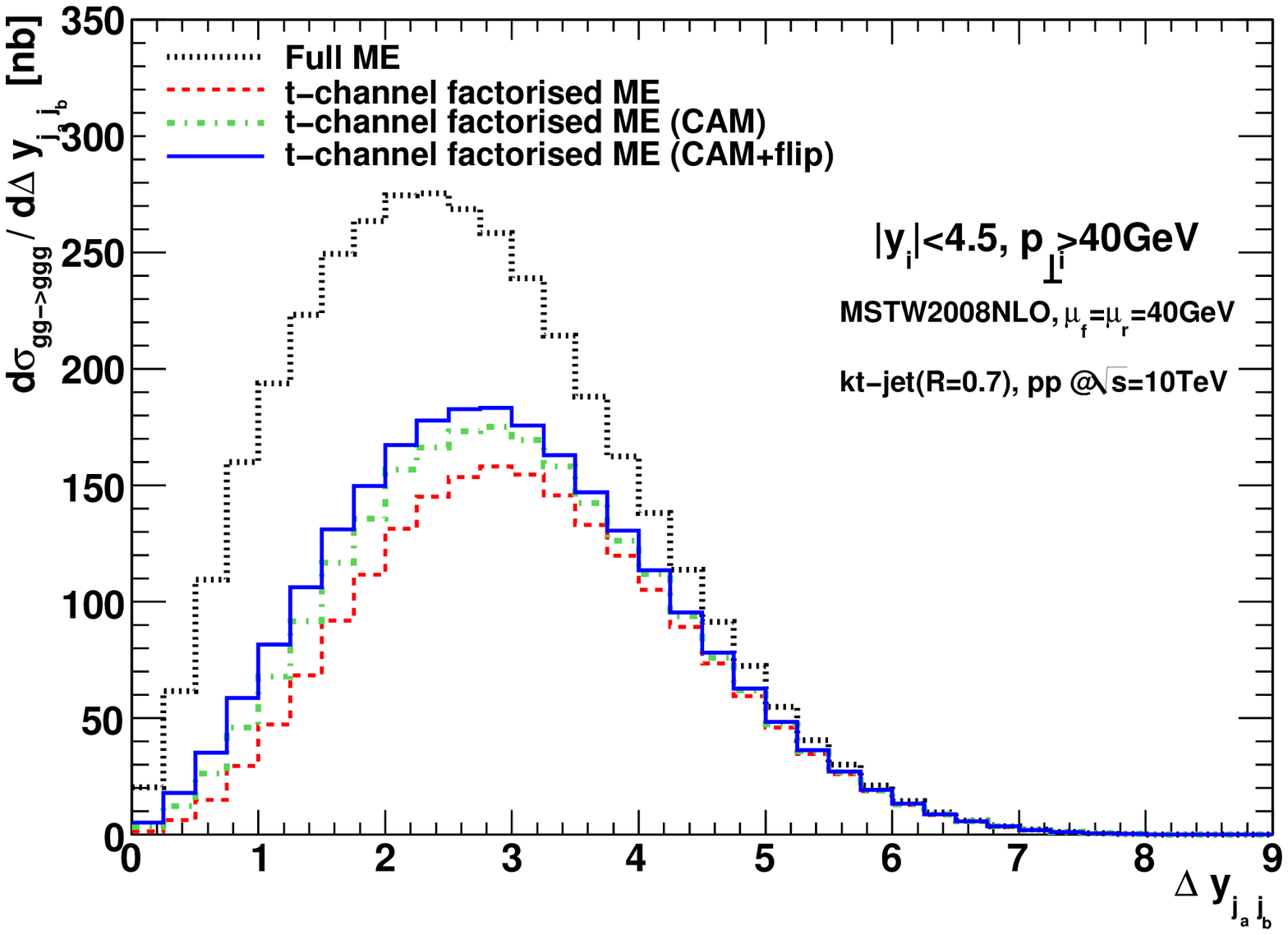}
  \epsfig{width=0.49\textwidth,file=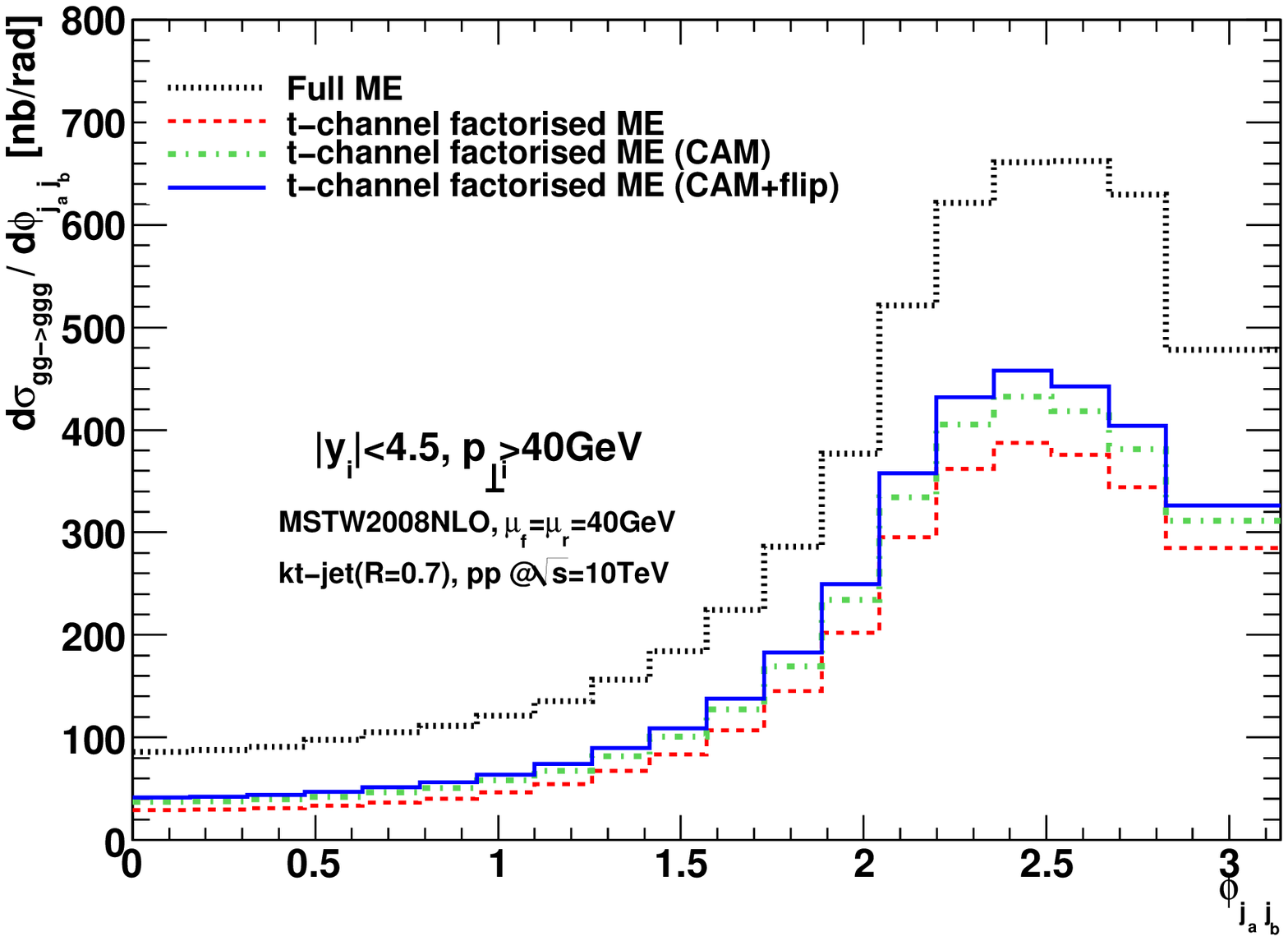}\\
  (e) \hspace{7.2cm}(f)\hspace{0.1cm}
  \caption{Results for d$\sigma$/d$\Delta y$ and d$\sigma$/d$\phi$ for $ud\to
    ugd$ (a)--(b), $ug\to ugg$ (c)--(d) and $gg\to ggg$ (e)--(f).  $\Delta y$
    is the rapidity difference between the most forward and most backward
    hard jet.  The black solid line represents the full matrix element, the
    red dashed line is the implementation based on the scattering of quark
    currents~\cite{Andersen:2009nu}, the blue dashed line is this result with
    the Colour Adjusted Multiplier (CAM) of Eq.~\eqref{eq:finalcol} and the
    green dashed line has the CAM and the effect of flipped helicities,
    Eq.~\eqref{eq:colsumandaverhelflipamp}.}
  \label{fig:3jres}
\end{figure}

\begin{figure}[!btp]
  \centering
  \epsfig{width=0.49\textwidth,file=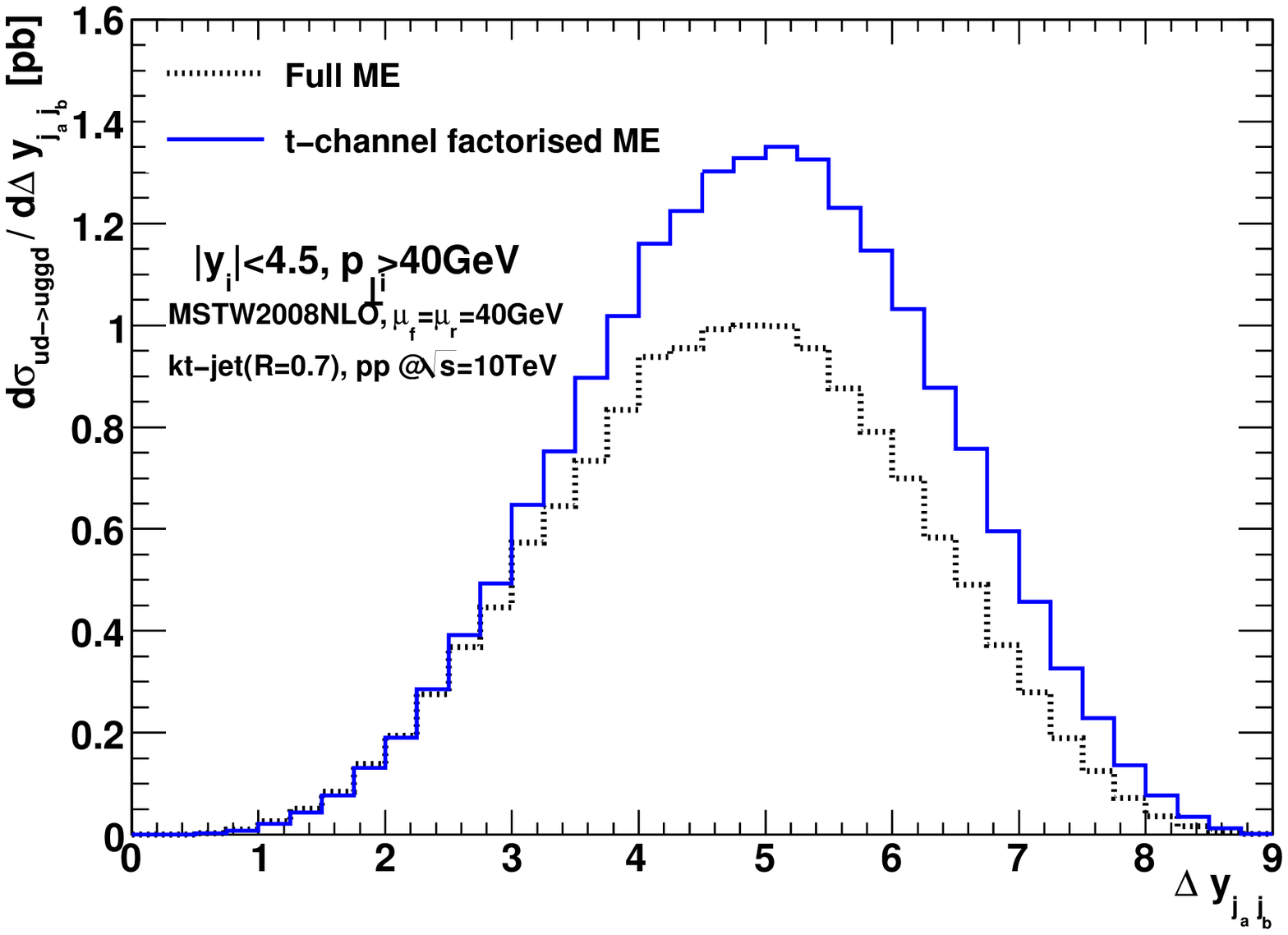}
  \epsfig{width=0.49\textwidth,file=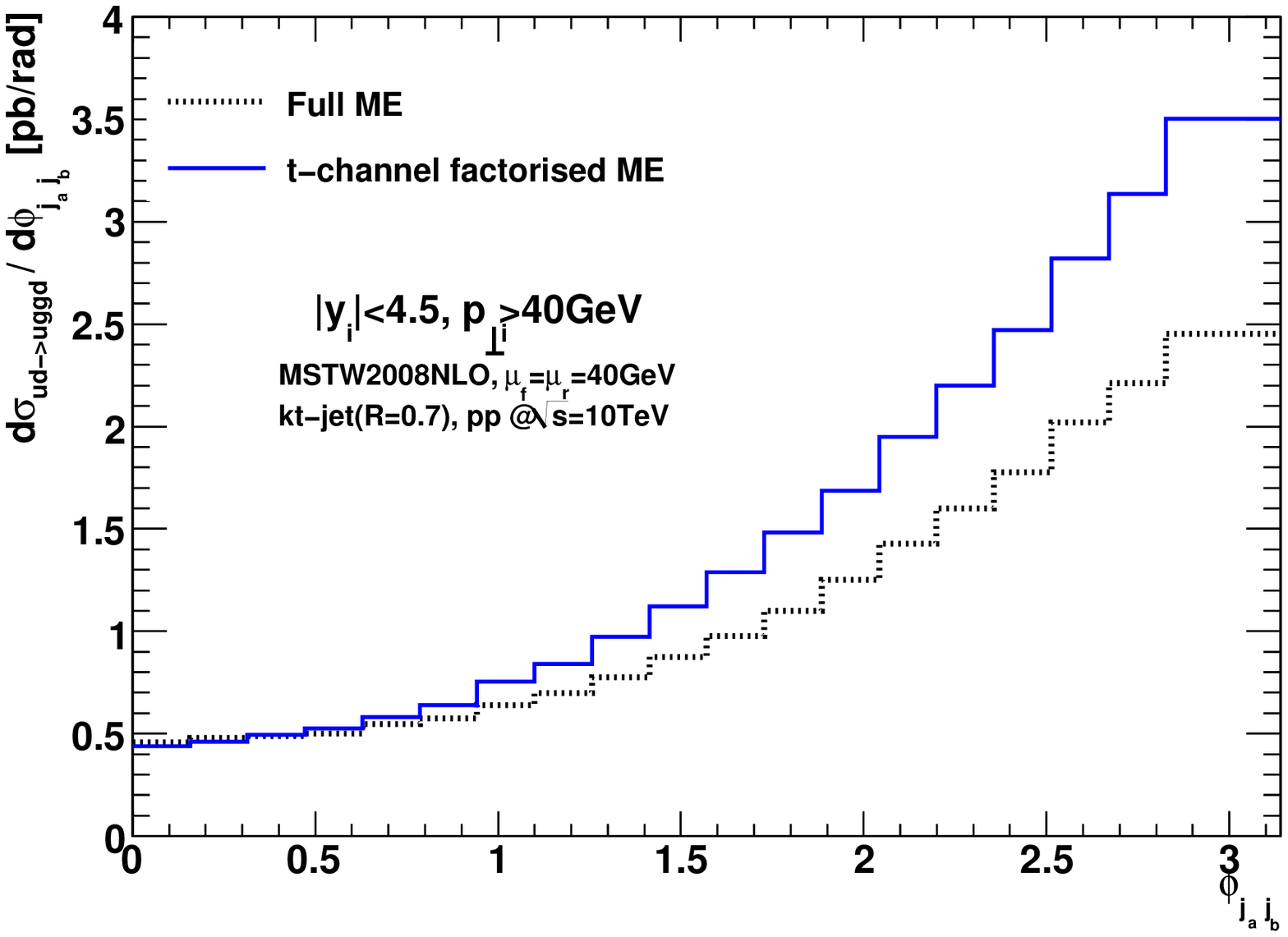} \\
  (a) \hspace{7.2cm}(b)\hspace{0.1cm}\\
  \epsfig{width=0.49\textwidth,file=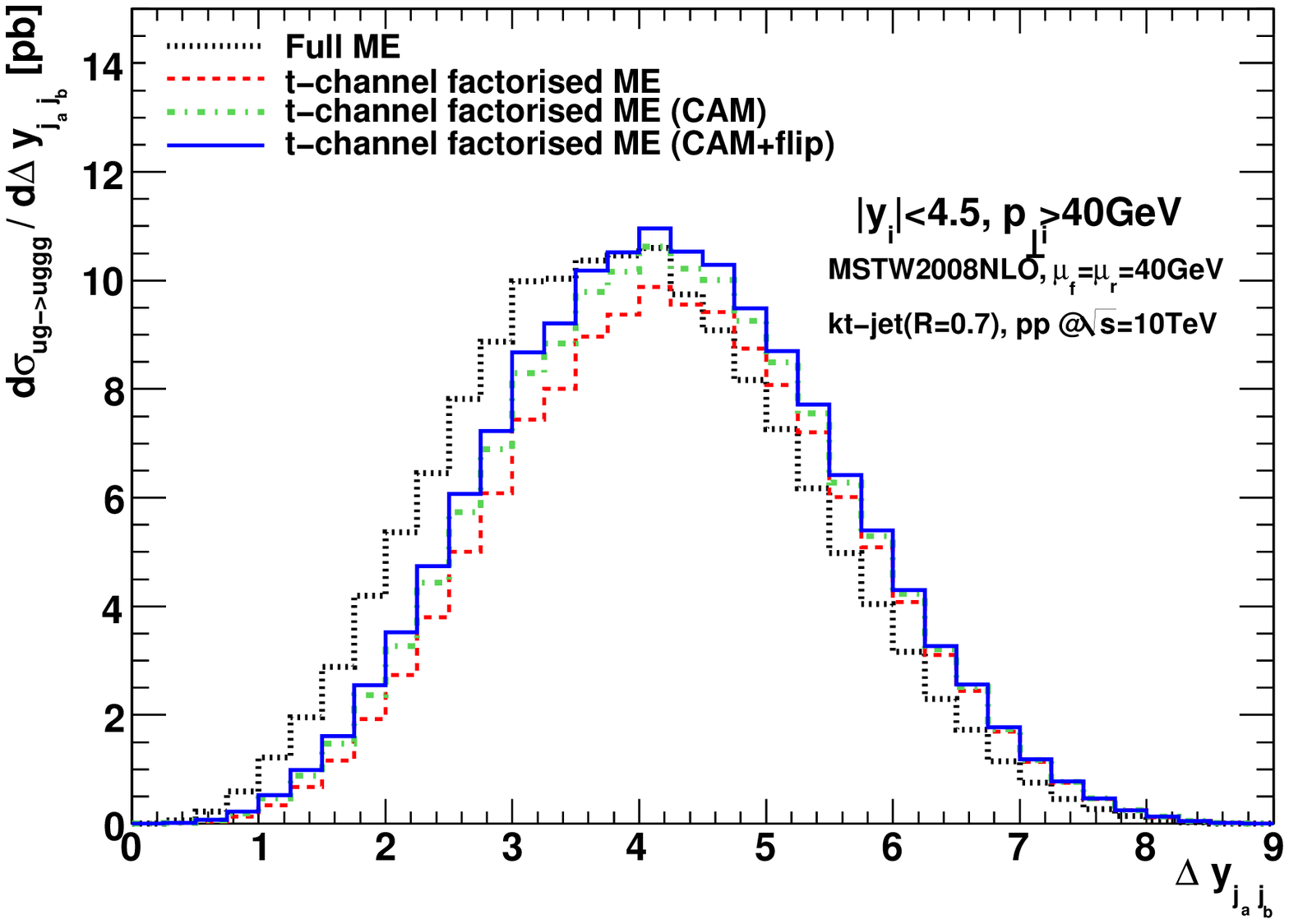}
  \epsfig{width=0.49\textwidth,file=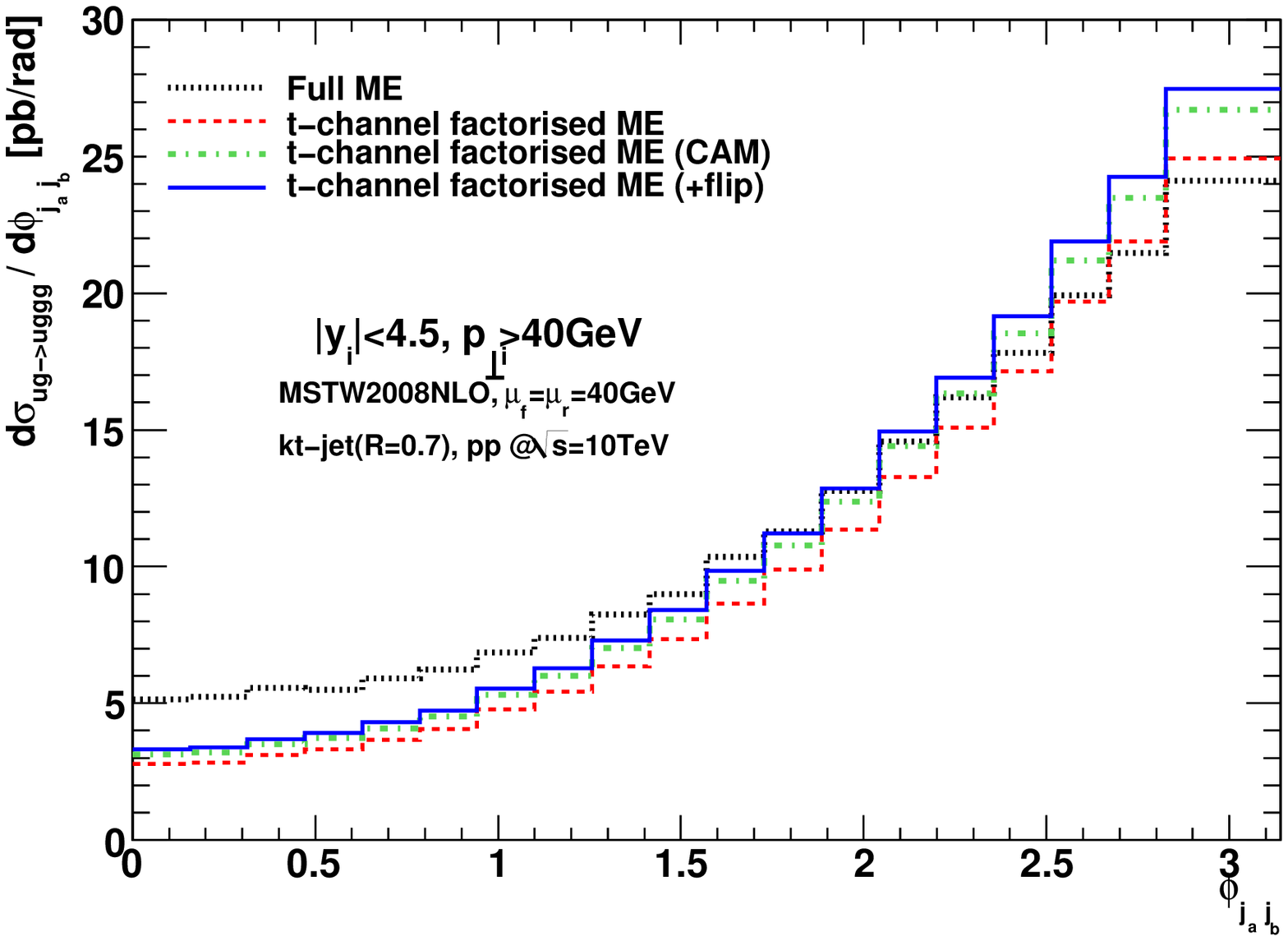} \\
  (c) \hspace{7.2cm}(d)\hspace{0.1cm}\\
  \epsfig{width=0.49\textwidth,file=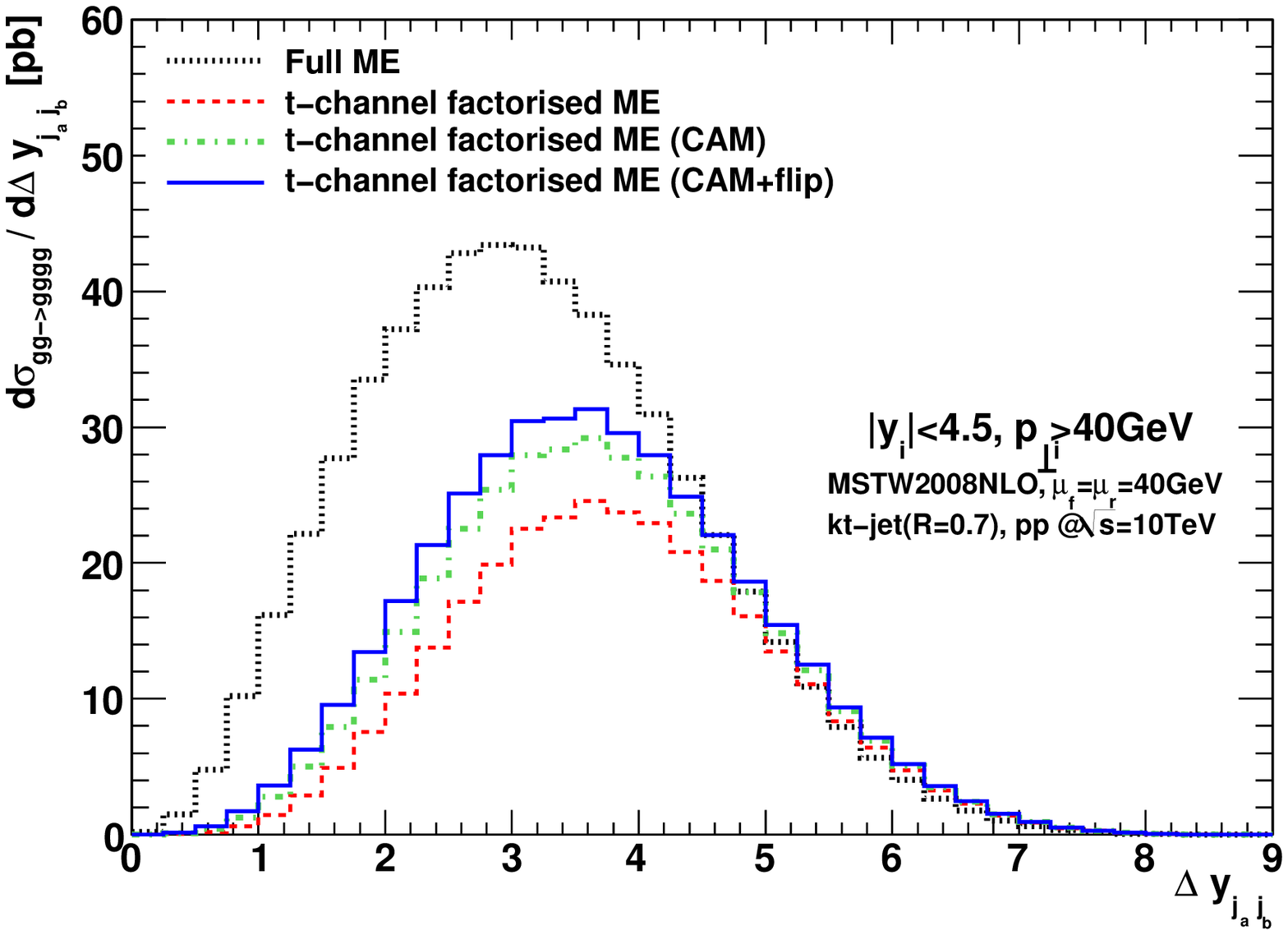}
  \epsfig{width=0.49\textwidth,file=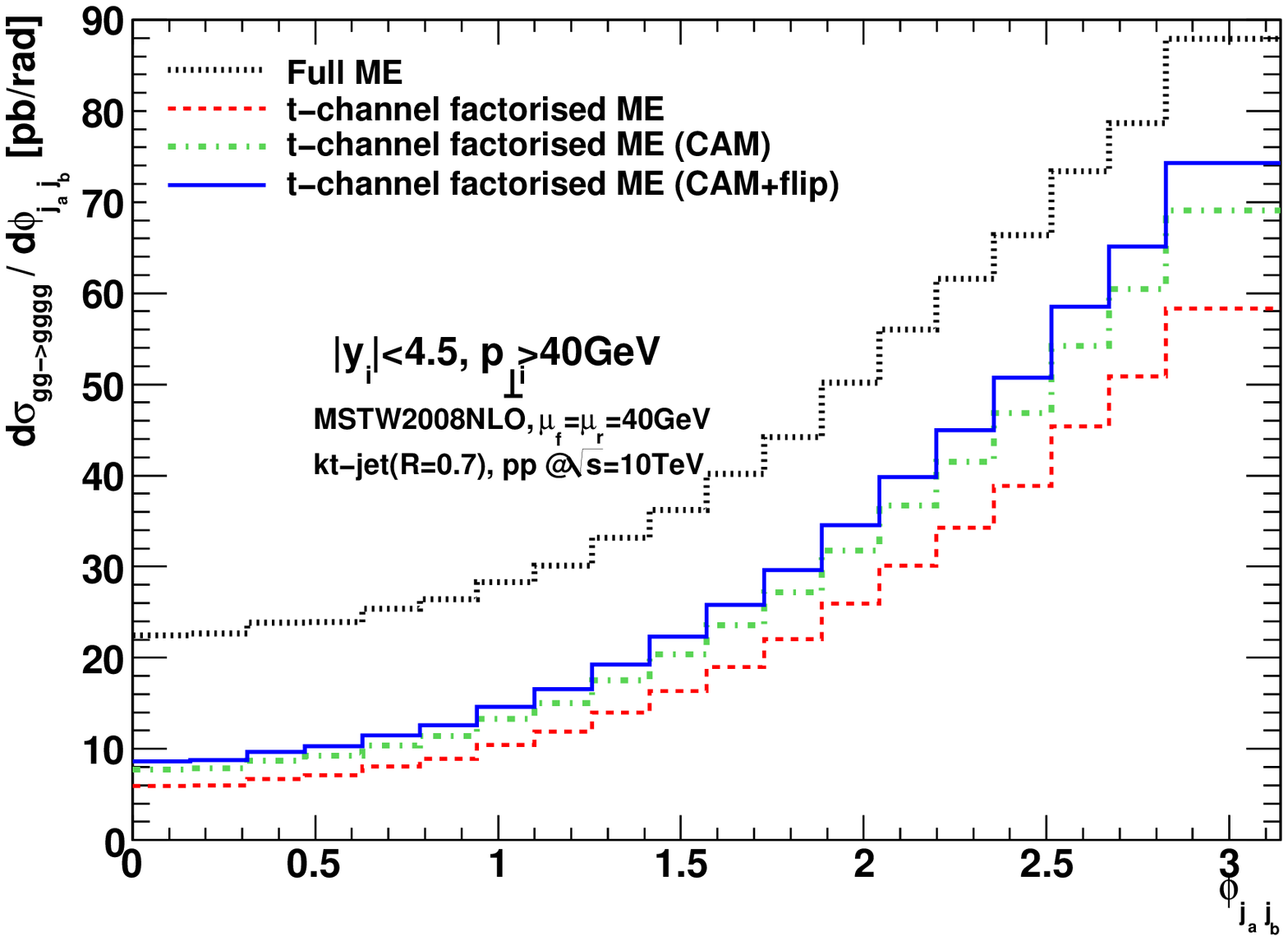}\\
  (e) \hspace{7.2cm}(f)\hspace{0.1cm}
  \caption{As in Fig.\ref{fig:3jres}, but now for the $4j$ final states: $ud\to uggd$
    (a)--(b), $ug\to uggg$ (c)--(d) and $gg\to gggg$ (e)--(f).}
  \label{fig:4jres}
\end{figure}

In this section, we show comparisons of our new treatment with the previous
treatment and the full matrix element, obtained from
Madgraph~\cite{Alwall:2007st}. We will concentrate on the changes introduced
in the approximations compared to the description in
Ref.\cite{Andersen:2009nu}, and will not show the results for just
2-jet-rates, since here the approximations are so good that the difference to
the full tree-level result is completely insignificant. Figures
\ref{fig:3jres} and \ref{fig:4jres} show the results for 3 and 4 jet final
states respectively (for both $qg$ and $gg$ initiated processes) within the
following cuts (identical to the ones used in Ref.\cite{Andersen:2009nu})
\begin{center}
  \begin{tabular}{|rl||rl|}
    \hline
    $p_{j_\perp}$& $> 40$ GeV & $|y_j|$ & $<$ 4.5\\\hline
  \end{tabular}
\end{center}
We show the differential cross section with respect to $\Delta y$, the
rapidity difference between the two jets extremal in rapidity, and $\phi$,
the angle in the transverse plane between these outer jets.  These are just
examples to illustrate the accuracy obtained in the perturbative
approximations. There is obviously no change compared to
Ref.\cite{Andersen:2009nu} in the cases of quark-quark-initiated processes,
which are just included here for completeness.


One can see that the effect of multiplying by the adjusted colour factor,
Eq.~\eqref{eq:finalcol}, alone (green lines, marked CAM) gives an improvement in all
cases.  It has a greater effect in the $4j$ cases compared to $3j$ cases, which agrees
with the interpretation of it as a contribution from the acceleration of the gluon.  One
would expect this to be greater when an extra jet is produced, and we do indeed see a
greater effect.  We then see a further, more modest, improvement when the channels where
the helicity of one of the gluons changes are also incorporated through
Eq.~\eqref{eq:colsumandaverhelflipamp}. The blue solid line in the plots is the sum total
of improvements, and are obtained within a formalism which, according to the results of
Ref.\cite{Andersen:2009nu}, is sufficiently simple that all orders in the perturbative
series can be summed directly.  We didn't go to higher than 4 jet final states here
because of the time it would take for the full matrix element results.  We were not
limited by the time for our formalism; the 4 jet results took about 5 minutes on a single
computer.


\section{Conclusions}
\label{sec:conclusions}
By exploring the scattering of specific helicity states in quark-gluon
scattering at tree level we have shown explicitly that the $t$-channel pole
can be described exactly as a contraction of two local currents. Furthermore,
we demonstrate that out of eight non-zero helicity possibilities, only two
suppressed possibilities have contributions that are not pure $t$-channel
poles. We extract a gauge-invariant definition for the $t$-channel current
generated by the scattering of a gluon. This at the same time directly proves
the assertions on the generality of quark and gluon scattering in the
Multi-Regge kinematic (MRK) limit made in Ref.\cite{Andersen:2009nu}, and offers
slight improvements in the description of scattering amplitudes in the
sub-asymptotic region. The formalism developed here is immediately applicable
in the resummation programme developed on the basis of
Ref.\cite{Andersen:2009nu} for the description of production of pure
multi-jets, and multiple jets in association with a $W,Z$ or $H$-boson.

\subsection*{Acknowledgements}
\label{sec:acknowledgements}
JMS would like to thank the CERN theory group for kind hospitality and is
supported by the UK Science and Technology Facilities Council (STFC). This
work was supported by the EC Marie-Curie Research Training Network ``Tools
and Precision Calculations for Physics Discoveries at Colliders'' under
contract MRTN-CT-2006-035505.


\appendix
\section{Spinor Representation}
\label{sec:spin-repr}

We use the following representation for the spinors.  For outgoing particles with
4-momentum $p$, $p^\pm=E\pm p_z$ and $p_\perp=p_x+ip_y$, we use
\begin{align}
  \label{eq:outspin}
  u^+(p)=\left(
    \begin{array}{c} \sqrt{p^+} \\ \sqrt{p^-}\ \frac{p_\perp}{|p_\perp|} \\ 0 \\
      0 \end{array} \right) \qquad \mathrm{and} \qquad u^-(p)=\left( \begin{array}{c} 0 \\
      0 \\ \sqrt{p^-} \
        \frac{p_\perp^*}{|p_\perp|}\\ -\sqrt{p^+} \end{array} \right).
\end{align}
For incoming particles with 4-momentum $p$ moving in the + direction, we use:
\begin{align}
  \label{eq:inspinp}
  u^+(p)=\left(
    \begin{array}{c} \sqrt{p^+} \\ 0 \\ 0 \\ 0 \end{array}
    \right) \qquad \mathrm{and} \qquad u^-(p)=\left( \begin{array}{c} 0 \\ 0 \\ 0\\
        -\sqrt{p^+} \end{array} \right).
\end{align}
For incoming particles with 4-momentum $p$ moving in the - direction, we use:
\begin{align}
  \label{eq:inspinm}
  u^+(p)=\left(
    \begin{array}{c} 0 \\ -\sqrt{p^-} \\ 0 \\ 0 \end{array}
    \right) \qquad \mathrm{and} \qquad u^-(p)=\left( \begin{array}{c} 0 \\ 0 \\
        -\sqrt{p^-} \\ 0\end{array} \right).
\end{align}
We use the following representation for the gamma matrices:
\begin{align}
  \label{eq:gammas}
  \begin{split}
    &\gamma^0 = \left(
      \begin{array}{cccc}
        0 & 0 & 1 & 0 \\ 0 & 0 & 0 & 1 \\ 1 & 0 & 0 & 0 \\ 0 & 1 & 0 & 0 
      \end{array} \right),\quad
    \gamma^1 = \left(
      \begin{array}{cccc}
        0 & 0 & 0 & -1 \\ 0 & 0 & -1 & 0 \\ 0 & 1 & 0 & 0 \\ 1 & 0 & 0 & 0 
      \end{array} \right),\ \\
    &\gamma^2 = \left(
      \begin{array}{cccc}
        0 & 0 & 0 & i \\ 0 & 0 & -i &  \\ 0 & -i & 0 & 0 \\ i & 0 & 0 & 0 
      \end{array} \right),\quad
    \gamma^3 = \left(
      \begin{array}{cccc}
        0 & 0 & -1 & 0 \\ 0 & 0 & 0 & 1 \\ 1 & 0 & 0 & 0 \\ 0 & -1 & 0 & 0 
      \end{array} \right).
  \end{split}
\end{align}


\bibliographystyle{JHEP}
\bibliography{gluonpapers}

\end{document}